# A State-of-the-art Survey on IDS for Mobile Ad-Hoc Networks and Wireless Mesh Networks

Novarun Deb[1], Manali Chakraborty[2], Nabendu Chaki[3]

Department of Computer Science & Engineering, University of Calcutta, India
92 APC Road, Kolkata 700009, India

[1]novarun.db@gmail.com, [2]manali4mkolkata@gmail.com, [3]nabendu@ieee.org

**Abstract.** *An Intrusion Detection System (IDS) detects malicious and selfish nodes in a network. Ad hoc networks are often secured by using either intrusion detection or by secure routing. Designing efficient IDS for wireless ad-hoc networks that would not affect the performance of the network significantly is indeed a challenging task. Arguably, the most common thing in a review paper in the domain of wireless networks is to compare the performances of different solutions using simulation results. However, variance in multiple configuration aspects including that due to different underlying routing protocols, makes the task of simulation based comparative evaluation of IDS solutions somewhat unrealistic. In stead, the authors have followed an analytic approach to identify the gaps in the existing IDS solutions for MANETs and wireless mesh networks. The paper aims to ease the job of a new researcher by exposing him to the state of the art research issues on IDS. Nearly 80% of the works cited in this paper are published with in last 3 to 4 years.*

## 1 Introduction

An intrusion may be defined as any action that attempt to compromise the integrity, confidentiality or availability of a resource or that goes against the security goals of a resource. This can be something as severe as stealing confidential data or misusing the email system for spam. External intrusion attempts are targeted to cause congestion, propagate incorrect routing information, prevent services from working properly or shutdown them completely. The internal intrusions could be a lot more damaging since malicious insider already belongs to the network as an authorized party. Since prevention of intrusions is not always possible, supportive intrusion detection techniques are required. Intrusion detection systems (IDSs) are not to prevent or deter attacks. Instead, the purpose is to alert the users about possible attacks, ideally in time to stop the attack or mitigate the damage [1].

Detecting Intrusion is difficult, particularly in the wireless domain. IDS often attempts to differentiate abnormal activities from the normal ones. Unfortunately, normal activities can be varied, and an attack may have resemblance to normal activities. Also, consistency of data in the time domain can detect unusual behavior but unusual behavior is not necessarily malicious. An IDS reaches perfection if it accurately detects majority of attacks and hardly makes any false or phantom detection. One basic assumption while designing any IDS should be that the attacker is intelligent and that the attacker has no shortage of resources.

An IDS essentially consists of three functions. First, the IDS must monitor some event and maintain the history of data related to that event. Second, the IDS must be equipped with an analysis engine that processes the collected data. It detects unusual or malicious signs in the data by measuring the consistency of data in the time domain. Currently there are two basic approaches to analysis: misuse detection and anomaly-based detection. Third, the IDS must generate a response, which is typically an alert to system administrators. It is up to the system administrator, how he wants to scrutinize the system after receiving an alert.

### 1.1 Why IDS solutions need to be different for MANET and WMN?

In MANETs, mobile nodes communicate with each other without the assistance of any infrastructure. The communication between nodes, not directly in transmission range, is performed via multiple hops, i.e., nodes cooperate and forward packets for other nodes. In addition to that, in WMN some nodes are stationary forming a kind of backbone and possibly functioning as gateway to further networks like the Internet. Thus from the architectural point of view, a MANET is necessarily a infrastructure-less or ad-hoc network, whereas a mesh networks uses a backbone.

Due to this basic difference in architecture, security issues, as considered for MANETs, are often quite different compared to that for WMNs. Some of these at times are in favor of the attacker and some in favor of protecting the network from intrusion. As for instance, let's consider that an attacker wants to launch a wormhole attack in both types of networks. When the mobility of the nodes is high in a MANET, it becomes practically impossible for the attacker to establish the "tunnel" through which packets are routed to another point in the network. The scenario is different in case of WMNs as the backbone routers are static and if such nodes are compromised, "tunnels" can be easily built through them. On the other hand, in case of a WMN, one may deploy more robust IDS solutions that uses the backbone of the mesh network. Thus, even if some of the protocols do well towards securing both MANET and WMNs, tailor-made solutions are required keeping in mind the differences of the two types of wireless networks. The study in this paper reveals that many gaps still exist for detecting intrusions, particularly in case of mesh networks, that has been relatively new and deployed more recently.

### 1.2 Organization of the Paper

In this paper, we have studied most recent works for IDS for MANETs and wireless mesh networks. In section 2 of this paper, we have reported and analyzed seven different IDS approaches for MANET out of which four has been published in last 4 years. In section 3, 100% of the six reported IDS approaches on wireless mesh networks have been proposed in last 2 years. Each of the sections 2 and 3 ends with separate tables highlighting the basic features and limitations of the existing IDS solutions.

Survey papers like this one often include simulation results to compare different approaches. However, here the authors have carefully avoided simulation for performance evaluation for a couple of reasons. Firstly, different approaches for intrusion detection assume different configurations in the network. Even the underlying routing protocols are not the same. Some of the approaches claim to be compatible with multiple existing routing protocols. However, there would be significant impact in the simulation results for such variance. This in turn would spoil the entire purpose of the simulation. Besides, the paper covers a total of 13 IDS solutions, most of which have been published very recently. Usually simulation based graphs are good for comparing a small number of alternate solutions. Thus, in stead of simulations, the authors have followed a careful analytic approach to compare the works referred.

## 2   IDS for Mobile Ad-hoc Networks

A Mobile Ad hoc Network (MANET) can be defined as a collection of mobile nodes that are geographically distributed and communicate with one another over a wireless medium. Ideally, in a MANET, each node is assumed to be a friendly node and participates willingly in relaying messages to their ultimate destinations. A mobile ad hoc network is built on ad-hoc demand and consists of some wireless nodes moving within a geographically distributed area. These nodes can join or leave the network at any time. MANET does not use fixed infrastructure and does not have a centralized administration. The nodes communicate on a peer-to-peer basis. The networks are built on the basis of mutual cooperation and trust. This leads to an inherent weakness of security.

Security in mobile wireless ad hoc networks was particularly difficult to achieve, notably because of the vulnerability of the links, the limited physical protection of each of the nodes, the sporadic nature of connectivity, the dynamically changing topology, the absence of a certification authority, and the lack of a centralized monitoring or management point [11]. This, in effect, underscored the need for intrusion detection, prevention, and related countermeasures. Like any other research area, one needs to do a systematic re-search of the existing works in the area of intrusion detection too. In a very recent paper [4], a number of IDS methods have been described for MANET. Although the compilation is good, no serious attempt has been initiated to identify the gaps in the works cited. Survey papers on IDSs for Wireless Mesh Networks are very few in numbers. In [2], contrary to the promise of the title of the paper, the methods referred are mostly applicable for wireless ad-hoc networks and MANETs.

Before one attempts to detect an intrusion, it is important to understand the nature and variation of attacks. The work by Martin Antonio [5] provides a fairly good analysis of MANET specific attacks and risk analysis by identifying assets, vulnerabilities and threats, usable for future MANET deployments and security work. Consequently, security solutions with static configuration would not suffice, and the assumption that all nodes can be bootstrapped with the credentials of all other nodes would be unrealistic for a wide range of MANET applications [3]. In practice, it is not possible to build a completely secure MANET system in spite of using the most complex cryptographic technique or so-called secured routing protocols. Some of the IDS algorithms that have been developed for MANETs are explained below. A comparative study is provided at the end of this section.

IDSX [1] was a cluster-based solution which used an extended architecture. The proposed solution acted as a second line of defense. Individual nodes could implement any IDS solution. IDSX was compatible with any IDS solution acting as the first line of defense. Simulation results show that the IDSX solution hardly produced any false positives. This was because it formed a consensus of the responses from different individual IDS solutions implemented in the nodes. Anomaly-based intrusion detection schemes could be deployed as the first line of defense. The proposed approach in [1] works within preset boundaries. In general, these are quite feasible and practical enough considering the nature of ad hoc networks. However, some of these may also be considered as the limiting constraints. IDSX has not been compared with any of the existing IDS solutions. Also, the proposed two-step approach would make the task of intrusion detection expensive in terms of energy and resource consumption.

In another innovative approach in [7], a solution is proposed using the concept of unsupervised learning in Artificial Neural Networks using Self-Organizing Maps. The technique named eSOM used a data structure called U-matrix which was used to represent data classes. Those regions which represented malicious information were watermarked using the Block-Wise method. Regions representing the benign data class was marked using the Lattice method. When a new attack is launched it causes changes in the pixel values. eSOM and the Watermarking technique can together identify if any pixel has been modified. This makes it very sensitive towards detecting intrusions. The authors claim that the solution is 80% efficient and remains consistent even with variations in mobility. Mentioned below are some of the drawbacks of this work [7]. The IDS employing eSOM would be trained in regular time periods. This results in additional overhead and takes a toll on the energy efficiency of the algorithm. However, the proposed intrusion detection engine has not been employed on various routing protocols for the detection of various types of attacks.

A leader election model for IDS in MANET based on the Vicky, Clarke and Groves (VCG) model was suggested in [8]. This requires every node to be as honest as possible. Leaders are elected in a manner which results in optimal resource utilization. Leaders are positively rewarded for participating honestly in the election process. By balancing the resource consumption amongst the nodes, a higher effective lifetime of the nodes was achieved. Experimental results indicate that the VCG model performs well during leader election by producing a higher percentage of alive nodes. However, the simulation results indicate that the normal nodes will carry out more duty of intrusion detection and die faster when there are more selfish nodes. Besides, as selfish nodes do not exhaust energy to run the IDS service, the percentage of packet analysis decreases with time. This is a severe security concern. In the case of static scenarios, the model elects the same node as leader repeatedly. This causes the normal nodes to die very fast.

CONFIDANT, another approach, similar to Watchdog and Path-rater scheme, has been proposed to overcome the drawbacks of the Watchdog and Path-rater by ignoring misbehaving nodes in the routing process [9]. Every node identifies its neighbors as friends and enemies, based on trust. Friends are informed of enemies. CONFIDANT claims that the packet delivery ratio is very high (97% and above). A couple of the issues that still leaves a gap in [9] are mentioned below. However, CONFIDANT keeps the packet delivery ratio high even in a very hostile environment, with the assumption that enough redundant paths are available to reach the destination node, bypassing the malicious ones. This assumption may not always hold. Also, in comparing the throughput of clients and servers, the CONFIDANT fortified network performs very poorly in contrast to the benign network.

SCAN [10] is based on two central ideas. First, each node monitors its neighbors for routing or packet forwarding misbehavior, independently. Second, every node observes its neighbors by cross validating the overhead traffic with other nodes. Nodes are declared malicious by a majority decision. This assumes that the network density is sufficiently high. However, in SCAN the network services are temporarily halted during intrusion detection. The lack of mobility reduces the detection efficiency. The assumption that network density is high may not always hold. Increase in mobility results in higher false positives. Besides, the packet delivery ratio can be heavily affected in the interval during which an attack is launched and when it is detected. Also, the communication overhead for SCAN grows with increase in the percentage of malicious nodes and with mobility.

In HIDS [3], another approach to the IDS has been proposed. HIDS is based on trust or reputation or honesty values of the mobile nodes. The trust value of a node is dynamically increased or decreased depending on its behavior. When a node behaves normally, it is positively rewarded; malicious activity results in negative rewards for that node. The trust on a node is recomputed based on its current honesty rate, and the rewards that it has earned. A comparative study between SCAN and HIDS shows that the latter involves lower storage and communicational overhead than SCAN. HIDS is inherently protected against false positives. However, maintaining up-to-date tables at different nodes, as required by HIDS, may not be an energy-efficient strategy. Also the proposed HIDS offers only a generic architecture for secure route detection. More detailed testing is required before it can be used for secure routing in MANET applications.

In [16] OCEAN was proposed as another extension to the DSR protocol. OCEAN also uses a monitoring system and a reputation system. The proposed solution exchanges second-hand reputation messages. OCEAN implements a stand-alone architecture to avoid phantom intrusion detections. Depending on whether a node participates in the route discovery process, OCEAN can detect misbehaving nodes and selfish nodes. However, the detection efficiency of OCEAN rapidly decreases with increase in the density of misbehaving nodes. Simulation results show that at high threshold values, other second hand protocols perform better with high mobility of the nodes. Also, the mobility model simulated for OCEAN is not very realistic. At high mobility, OCEAN is very sensitive to change of the threshold parameter, while second hand protocols are more consistent over varying threshold limits. OCEAN is not quite effective in penalizing misleading nodes.

A hybrid solution, proposed in [17], combines the Watchdog and Path-Raters scheme proposed by Marti et al. and SCAN[10]. However, neither SCAN nor Watchdog and Path-raters address the mobility issue that well. As a result, this hybrid solution also suffers from the same problems. Besides, there are no fixed nodes which can behave as umpires. There must be some kind of a leader election model which runs in every node to select the Umpire nodes. This results in an increased overhead and energy consumption. The authors did mention the scenario where Umpire nodes themselves can become malicious. However, it still remains as a drawback of the method. In order to detect DoS attacks like flooding, the criteria for attack detection cannot be so rigid. Also, the history of a node that had being behaving normal, should be taken in to consideration before writing it off as malicious as soon as it deviates from normal behavior.

Table 1. Summary on Comparison for Different IDS for Mobile Ad-hoc Networks

| IDS Reference | Under-lying Routing Protocol | Architecture | Types of attacks addressed | Comments |
|---|---|---|---|---|
| IDSX [1] (2007). | Compatible with any routing protocol | Extended Hierarchical Architecture | Routing misbehavior - dirty packet forwarding. | 1. The solution talks about a two-step approach. This leads to making the intrusion detection approach quite expensive in terms of energy and resource consumption. |
| Neural Networks and Watermarking Technique [7] (2007) | AODV | Self Organizing Maps (Neural Networks) | Routing behavior attack and Resource utilization attack. | 1. The IDS using eSOM needs to be trained in regular interval. This additional overhead affects the energy efficiency of the algorithm. 2. The proposed intrusion detection engine has not been employed for various routing protocols for detection of different attacks. |
| CONFIDANT [9] (2002) | DSR | Distributed and Cooperative. | Packet drop attack. | 1. CONFIDANT assumes that there are enough nodes to provide harmless alternate partial paths around malicious nodes. This may not always hold. 2. A CONFIDANT fortified network with one third malicious nodes does not provide any additional benefits over a regular benign DSR network without malicious nodes. |
| HIDS [3] (2008) | Compatible with reactive And proactive routing protocols. | Distributed and Collaborative | Packet drops, black-hole attack, Resource utilization attacks | 1. Maintenance of tables at different nodes affects energy efficiency and communication overhead. 2. Detailed testing is required before it can be used for secure routing in MANET applications. |
| Leader Election Model [8] (2008) | Not specified. | Based on the Vickey, Clarke, and Groves (VCG) model by which truth-telling is the dominant strategy for each node. | Resource utilization attack-selfish nodes. | 1. Simulation results indicate that normal nodes will work more to detect intrusion and die faster in presence of selfish nodes. 2. As selfish nodes do not exhaust energy to run the IDS service, the percentage of packet analysis decreases with time. 3. In the case of static scenarios, the model elects the same node as leader repeatedly. This causes the normal nodes to die very fast. |
| SCAN [10] (2006) | AODV | Distributed and Collaborative | Routing misbehavior and packet forwarding misbehavior | 1. Network services are temporarily halted during intrusion detection.<br>2. Lack of mobility reduces the detection efficiency.<br>3. The assumption that network density is high may not always hold. Increase in mobility results in higher false positives.<br>4. Packet delivery ratio can be heavily affected in the interval between an attack is launched and when it is detected.<br>5. The communication overhead steadily increases with increase in the percentage of malicious nodes and with mobility. |
| OCEAN [16] (2003) | Not identified | Stand - alone IDS | Routing behavior attack, resource utilization attack, rushing attack. | 1. At high faulty thresholds, approaches like SEC-HAND protocols are able to perform better than OCEAN at high mobility.<br>2. At lower numbers of misbehaving nodes, the performance of OCEAN falls drastically.<br>3. OCEAN is not very effective in thwarting the throughput of the misleading nodes. |
| A System of Umpires [17] (2010) | Not identified | Stand - alone IDS for single user; Collaborative IDS for double and triple Users | Routing misbehavior attack and Packet Dropping attack. | 1. Umpires are not static. Some kind of leader election is required. This may require additional energy.<br>2. Attack detection criteria are very rigid.<br>3. Nodes are not rewarded for normal behavior. |

## 3. IDS for Wireless Mesh Networks

The proposed methodology successfully detects any moving object maintaining low computational complexity and low memory requirements.

Although mobility of nodes was removed and a certain infrastructure was established for Sensor Networks, yet these remained vulnerable to security threats. Researchers realized that mobility is a feature which cannot be compromised with as it provides tremendous flexibility to end users. Yet, retaining an infrastructure would definitely be helpful. All these underlying observations led to the conclusion that a different type of network must be designed which incorporates both the mobility of clients and a basic infrastructure. This had been a major driving factor behind the inception of Wireless Mesh Networks.

Wireless mesh networks (WMNs) consist of mesh routers and mesh clients, where mesh routers have minimal mobility and form the backbone of WMNs [2]. They provide network access for both mesh and conventional clients. The integration of WMNs with other networks such as the Internet, cellular, IEEE 802.11, IEEE 802.15, IEEE 802.16, sensor networks, etc., can be accomplished through the gateway and bridging functions in the mesh routers. WMNs include mesh routers forming an infrastructure for clients that connect to them. The WMN infrastructure/backbone can be built using various types of radio technologies.

The client meshing provides peer-to-peer networks among client devices. In this type of architecture, client nodes constitute the actual network to perform routing and configuration functionalities as well as providing end user applications to customers.

Hence, a mesh router is not required for these types of networks. In Client WMNs, a packet destined to a node in the network hops through multiple nodes to reach the destination. Client WMNs are usually formed using one type of radios on devices. Moreover, the requirements on end-user devices is increased when compared to infrastructure meshing, since, in Client WMNs, the end-users must perform additional functions such as routing and self-configuration.

Mesh clients can access the network through mesh routers as well as directly meshing with other mesh clients. While the infrastructure provides connectivity to other networks such as the Internet, Wi-Fi, WiMAX, cellular, and sensor networks; the routing capabilities of clients provide improved connectivity and coverage inside the WMN. The hybrid architecture will be the most applicable case in our opinion.

The redundancy and self-healing capabilities of WMNs provide for less downtime, with messages continuing to be delivered even when paths are blocked or broken. The self-configuring, self-tuning, self-healing, and self-monitoring capabilities of mesh can help to reduce the management burden for system administrators. Besides, the advanced mesh networking protocols coordinate the network so that nodes can go into sleep mode while inactive and then synchronize quickly for sending, receiving, and forwarding messages. This ability provides greatly extended battery life.

A mesh network can be deliberately *over-provisioned* simply by adding extra devices, so that each device has two or more paths for sending data. This is a much simpler and less expensive way of obtaining redundancy than is possible in most other types of networks. In comparison to the cost of point-to-point copper wiring and conduit required for traditional wired networks, wireless mesh networks are typically much less expensive. The protocols that have been developed so far for WMNs are described briefly. A comparative study is provided at the end of this section.

A technique was devised based on the communication history between two communicating clients through a common set of routers in [15]. Individual trust relationships are evaluated for both clients sharing the common set of routers. Malicious clients are detected based on threshold values. The algorithm performs well when the density of malicious nodes is low. Routers in the path have to perform $O(N^2)$ operations to cooperatively reach a conclusion. It is found that false positives are reduced to a great extent but not eliminated. The algorithm performs better only when the percentage of misbehaving clients is smaller. Performance degrades as malicious activity within the network increases.

RADAR [12] introduces a general concept of reputation. Highly detailed evaluation metrics are used to measure the behavior of mesh nodes. This allows RADAR to better classify / distinguish normal behavior from anomalous activity. RADAR takes into consideration the spatio-temporal behavior of nodes before declaring them as malicious. Simulation results show that RADAR detects routing loops with higher false alarms. The algorithm is resilient to malicious collectives for subverting reputations; but involves a relatively high latency for detection of DoS attacks. The Detection Overhead Ratio (DOR) is directly proportional to the number of anomaly detectors and the size of detection window implemented in the algorithm.

Although developed initially for wired networks, Principal Component Analysis (PCA) based method [11] could also be implemented for wireless networks. The threshold value used in [11] for detecting malicious nodes assumes that network traffic follows the normal distribution. Tuning the threshold also reduces the number of phantom intrusion detections considerably. The proposed solution is energy-efficient. However, despite the promises, the PCA based method in [11] is not consistent to variations in normal network traffic due to unrealistic assumptions in the method. Anomalies such as node outages cannot be detected as this method [11] looks for spurious traffic generation. A statistical analysis of how the behavior varies with changing threshold values is yet to be performed.

In [14] a solution to defend against selective forwarding attacks based on AODV routing protocol is presented. The algorithm works in two phases – detecting malicious activities in the network and identifying the attacker, respectively. However, the proposed methodology of [14] suffers from some serious limitations. The proposed scheme fails to detect attackers when the threshold value is less than the throughput. Even in the absence of an attacker, the throughput is low when the detection threshold is higher than throughput of the path. Performance overhead of the system increases with increase in the density of malicious nodes.

OpenLIDS [13] analyzes the ability of mesh nodes to perform intrusion detection. Due to the resource constraints of mesh nodes, detailed traffic analysis is not feasible in WMNs. An energy – efficient scheme was proposed in OpenLIDS. Results show that performance improved for detecting malicious behavior in mesh nodes. OpenLIDS is an improvement over other signature-based approaches both in terms of memory requirements and packet delivery ratio. However, simulation results show that OpenLIDS is unable to distinguish an RTP stream from a UDP DoS flood with fixed source and destination ports. For new connections, this approach is not as efficient as expected as generating and receiving connection tracking events is costly.

In [6], a framework has been proposed that is based on a reputation system. This isolates ill-behaved nodes by rating their reputation as low, and distributed agents based on unsupervised learning algorithms, that are able to detect deviations from the normal behavior. The solution is very effective in detecting novel intrusions. This algorithm had already been deployed for WSNs. Experimental results show that even though redundancy reduces drastically in WMNs the proposed method works efficiently. However, the approach is not fast enough to prevent the neighbor nodes from being affected by an attack. Also, initially the solution [6] cannot exactly determine the source of the anomaly. Therefore, the system reduces the reputation of all the nodes within the malicious region.

**Table 2.** Summary on Comparison for Different IDS for Wireless Mesh Networks

| IDS Reference | Under-lying Routing Protocol | Architecture | Types of attacks addressed | Comments |
|---|---|---|---|---|
| Trust based approach I [14] (2008). | AODV | Distributed System | Gray hole attacks. | 1. The overhead of the system increases with the number of attackers. 2. When detection threshold is less than the throughput of a path, attacks will not be detected and network throughput will suffer. 3. On the contrary, when the detection threshold is higher than throughput of the path, the throughput would suffer even if there is no attacker. |

| Trust based approach II [15] (2008) | Not specified. | Distributed Systems | Misbehavior of a node | 1. The detection efficiency decreases and false positive rate increases with the increase of percentage of malicious clients.<br>2. False positives are reduced to a great extent but not eliminated. |
|---|---|---|---|---|
| Principal Component Analysis (PCA) [11] (2008). | Not specified | Distributed Systems | DoS, port scan, jamming etc. | 1. Anomalies such as node outages are not detected as the method looks for spurious traffic generation.<br>2. Analysis on performance evaluation with changing threshold values is yet to be performed.<br>3. The method is not consistent due to unrealistic assumptions on network traffic. |
| RADAR [12] (2008). | DSR | Distributed Systems | Malicious behavior of a node, DOS Attack, Routing Loop Attack. | 1. Higher false alarms.<br>2. Resilient to malicious collectives for subverting reputations.<br>3. High latency for detection of DoS attacks.<br>4. The Detection Overhead Ratio (DOR) is a linear overhead. |
| OpenLIDS [13] (2009). | Not specified | Distributed Systems | Resource starvation attacks, mass mailing of internet worms, IP spoofing. | 1. Higher false positives as OpenLIDS is unable to distinguish between RTP stream and a UDP DoS flood with fixed source and destination ports.<br>2. Not as efficient for new connections.<br>3. It is not possible to arbitrarily adjust timeout values. |
| Reputation systems and self-organizing maps. [6] (2010). | Not specified. | Distributed agent based Systems | Routing misbehavior and resource utilization attacks. | 1. It is assumed that the confidentiality and integrity cannot be preserved for any node.<br>2. The reputation system identifies the attacked node immediately. However, it is not fast enough to prevent the neighbor nodes from being affected |

## 4. Conclusion

The thirst of flexibility in operations and application requirements for wider access has triggered the evolution of ad-hoc networks. The infrastructure-less ad-hoc networks offer even greater flexibility when the nodes are mobile. This flexibility is however two-fold. Just the way, a greater number of applications are made possible in ad-hoc networks, especially for MANETs, the lack of centralized control, dedicated security infrastructure, non-standard topology, etc. offers additional "flexibility" to the intruder as well. Designing efficient IDS that would not affect the performance of the network is in fact an uphill task due to the vulnerability of the links, the limited physical protection of the nodes, the irregularity and dynamic changes in topology, and the lack of a centralized authority and monitoring. In spite of this, recent works propose adept IDS methodologies that extract the advantages of base station in sensor networks and the backbone in wireless mesh networks. In section 1 of the paper, the reasons for avoiding simulation have been explained. Instead, critical analytic comparisons are done for 13 different IDS solutions. The findings are summarized in tables I to II. An IDS needs a scalable architecture to collect sufficient evidences to detect those attacks effectively. Researchers are now being motivated to design a new IDS architecture that involves cross layer design to efficiently detect the abnormalities in the wireless networks. The selection of correct combination of layers in the design of cross layer IDS is very critical to detect attacks targeted at or sourced from any layers rapidly. It is optimal to incorporate MAC layer in the cross layer design for IDS to detect DoS attacks. This cross layer technique incorporating IDS leads to an escalating detection rate in the number of malicious behavior of nodes increasing the true positive and reducing false positives in the MANET. The current study may be extended to review recent works on cross-layer IDS architecture.